\documentclass{PoS}

\usepackage{epsf}
\usepackage{amssymb}
\usepackage{amsmath}
\usepackage{amsfonts}
\usepackage{cite}
\usepackage[small]{caption2}

\usepackage[T1]{fontenc}
\usepackage{psfrag,epsfig,graphicx,graphics}

\newcommand{\be}{\begin{equation}}
\newcommand{\beq}{\begin{equation}}
\newcommand{\ee}{\end{equation}}
\newcommand{\eq}{\end{equation}}
\newcommand{\eeq}{\end{equation}}

\newcommand{\bea}{\begin{eqnarray}}
\newcommand{\eea}{\end{eqnarray}}

\def\slashchar#1{\setbox0=\hbox{$#1$}
   \dimen0=\wd0
   \setbox1=\hbox{/} \dimen1=\wd1
   \ifdim\dimen0>\dimen1
      \rlap{\hbox to \dimen0{\hfil/\hfil}}
      #1
   \else
      \rlap{\hbox to \dimen1{\hfil$#1$\hfil}}
      /gdatdafinal2.tex
   \fi}

\newcommand{\fV}{f_{3\,\rho}^V}
\newcommand{\fA}{f_{3\,\rho}^A}

\title{Hard exclusive electroproduction of $\rho_T$ at twist 3}

\ShortTitle{Hard exclusive electroproduction of $\rho_T$ at twist 3}

\author{I.~V.~Anikin\\
Bogoliubov Laboratory of Theoretical Physics, JINR,
             141980 Dubna, Russia\\
E-mail: \email{anikin@theor.jinr.ru}}

\author{D.~Yu.~Ivanov\\
Sobolev Institute of Mathematics, 630090 Novossibirsk, Russia\\
E-mail: \email{d-ivanov@math.nsc.ru}}

\author{B.~Pire\\
CPHT, {\'E}cole Polytechnique, CNRS, 91128 Palaiseau Cedex, France\\
E-mail: \email{pire@cpht.polytechnique.fr}}

\author{L.~Szymanowski\\
Soltan Institute for Nuclear Studies, PL-00-681 Warsaw, Poland\\
E-mail: \email{Lech.Szymanowski@fuw.edu.pl}}

\author{\speaker{S.~Wallon}%\thanks{A footnote may follow.}
\\
        LPT, Universit{\'e} Paris-Sud, CNRS, 91405 Orsay, France \ {\em \&} \\
UPMC Univ. Paris 06, facult\'e de physique, 4 place Jussieu, 75252 Paris Cedex  05, France\\
        E-mail: \email{wallon@th.u-psud.fr}}

\abstract{
Exclusive processes in hard electroproduction  are one of the best place for understanding QCD in view of its factorization properties.
In particular, in the limit of
asymptotic $\gamma^* p$ center of mass energy, they give an access to 
the perturbative Regge properties of QCD. 
The HERA experiment recently provided precise data for $\rho$ electroproduction, including all
spin density matrix elements. From QCD, it is expected that such a process should factorize between
a hard (calculable) coefficient function, and hadronic ($p$ and $\rho$) matrix elements. Such
a factorization is up to now only proven for a longitudinaly polarized $\rho$.  Within the $k_t$-factorization
approach (valid at large  $s_{\gamma^* p}$),
we evaluate the impact factor of the transition $\gamma^{*} \to \rho_{T}$ taking into account
the twist $3$ contributions. We show that
a gauge invariant expression is obtained with the help of QCD equations of motion. 

More generally, relying on these equations and 
on the invariance under rotation on the light-cone of
 the factorized amplitude, the  non-perturbative Distribution Amplitudes can be reduced to a minimal set.
 This opens the way to a
consistent treatment of factorization for exclusive processes with a transversally polarized vector meson, and clarifies  the equivalence of two proposed parametrizations of the $\rho_{T}$ distribution amplitudes, illustrated here by the  $\gamma^{*} \to \rho_{T}$ impact factor.
}

\FullConference{XVIII International Workshop on Deep-Inelastic Scattering and Related Subjects, DIS 2010\\
		April 19-23, 2010\\
		Firenze, Italy}

\begin{document}

\section{Introduction}
\label{Sec_Int}

Studies of hard exclusive reactions rely on the factorization properties of the leading twist amplitudes \cite{fact}.
%for deeply virtual Compton scattering and deep exclusive meson production \cite{DGP}. 
The leading twist
distribution amplitude (DA) of a transversally polarized vector meson is
chiral-odd.
 Moreover it decouples in leading twist electroproduction amplitudes
even when another
chiral-odd quantity is involved \cite{DGP} unless in reactions with more than two final
hadrons \cite{IPST}.
Thus
transversally polarized $\rho-$meson production is generically governed by  
twist 3 contributions for which  a pure collinear
factorization fails  due to the appearance of end-point singularities \cite{MP,AT}.
The meson quark gluon structure within collinear factorization may be described by Distribution Amplitudes (DAs), %which have been classified in
classified in \cite{BB}. Measurements \cite{exp} of the 
$\rho_T-$meson production amplitude  in
 photo and electro-production   show that it
%Although non-dominant for deep electroproduction, this amplitude 
 is by no means negligible.
%  Up to now, experimental information comes from electroproduction on a proton
%   or nucleus. 
We  consider here the case of  very high energy collisions at colliders, for which
%, which was studied at HERA \cite{expHigh}.  
future progress  may come from real or virtual photon photon collisions \cite{IP,PSW}.
%, which could be accessible either at electron
%  positron colliders or in ultraperipheral collisions at hadronic colliders 
In the literature there are two approaches to the factorization of the
scattering amplitudes in exclusive processes at leading and higher twists. 
%The first approach \cite{APT,AT}
The Light-Cone Collinear Factorization (LCCF) \cite{APT,AT} extends the inclusive approach \cite{EFP}
%the Ellis-Furmanski-Petronzio (EFP) method  
 to exclusive processes, dealing with the factorization in the momentum space around the dominant light-cone
direction, while  the Covariant Collinear Factorization (CCF) approach
in coordinate space was succesfully applied in \cite{BB} for a systematic
description of DAs of hadrons carrying different
twists.  
% Although being quite
% different and using different DAs, both
% approaches can be applied to the description of the same processes, and the question whether they are equivalent naturally rises. 
We show  \cite{us} that these two descriptions are equivalent at twist 3. 
For that, we perform
our analysis  within LCCF method in momentum space and use
the invariance
of the scattering amplitude under rotation of the light-cone. 
% vector $n^\mu$,
%which we call
% $n$-independence condition.  
This method leads to introduce relevant soft correlators which are generally not independent ones, and are reduced  to a minimal  independent set  with the use of equation of motions and of the light-cone-fixing vector independence condition.  A dictionary is obtained between LCCF and CCF correlators, proving the equivalence between LCCF and CCF approaches.
We  illustrate this equivalence by   calculating
within both methods the impact factor $\gamma^* \to \rho_T$, 
%which is the building block of the
%description of the $\gamma^* \, p \to \rho \, p$ process at large $s\,,$ 
up to twist 3
accuracy. 
%We prove the full consistency between the two results. This as been shown in Ref.\cite{us}

\section{LCCF factorization of exclusive processes}
 \label{Sec_LCCF}

%\subsection{Factorization beyond leading twist}
% \label{SubSec_fact}

The amplitude for the exclusive process $A \to \rho \, B$ is,  in
 the momentum representation and in axial
gauge reads ($H$ and $H_\mu$ are  2- and 3-parton coefficient functions,
 respectively)
\begin{eqnarray}
\label{GenAmp}
{\cal A}=
\int d^4\ell \, {\rm tr} \biggl[ H(\ell) \, \Phi (\ell) \biggr]+
\int d^4\ell_1\, d^4\ell_2\, {\rm tr}\biggl[
H_\mu(\ell_1, \ell_2) \, \Phi^{\mu} (\ell_1, \ell_2) \biggr] + \ldots \,.
\end{eqnarray}
In (\ref{GenAmp}), the soft parts $\Phi$ are  the
Fourier-transformed 2- or 3-parton correlators which are matrix elements of non-local operators.
To factorize the amplitude, we  choose the dominant direction around which
we  decompose our relevant momenta and  we Taylor expand the hard part.
Let $p\sim p_\rho$ and $n$ be two light-cone vectors ($p \cdot n =1$).  Any vector $\ell$ is then expanded as
\begin{eqnarray}
\label{k}
\ell_{i\, \mu} = y_i\,p_\mu  + (\ell_i\cdot p)\, n_\mu + \ell^\perp_{i\,\mu} ,
\quad y_i=\ell_i\cdot n ,
\end{eqnarray}
and  the integration measure in (\ref{GenAmp}) is replaced as
$d^4 \ell_i \longrightarrow d^4 \ell_i \, dy_i \, \delta(y_i-\ell\cdot n) .$
The hard part  $H(\ell)$ is then expanded around
the dominant  $p$ direction:
\begin{eqnarray}
\label{expand}
H(\ell) = H(y p) + \frac{\partial H(\ell)}{\partial \ell_\alpha} \biggl|_{\ell=y p}\biggr. \,
(\ell-y\,p)_\alpha + \ldots
\end{eqnarray}
where $(\ell-y\,p)_\alpha \approx \ell^\perp_\alpha$ up to twist 3.
To obtain a factorized amplitude, one performs 
an
 integration 
 by parts
to replace  $\ell^\perp_\alpha$ by $\partial^\perp_\alpha$ acting on
the soft correlator.
 This leads to new operators containing
transverse derivatives, such as $\bar \psi \, \partial^\perp \psi $,
 thus requiring
additional DAs
$\Phi^\perp (l)$.
%This procedure accomplishes the factorization of the amplitude in momentum
%space. 
Factorization is then achieved by 
 Fierz decomposition on a set of relevant Dirac $\Gamma$ matrices, and we end up with
\bea
\label{GenAmpFac23}
\hspace{-.4cm}{\cal A}=
  {\rm tr} \left[ H_{q \bar{q}}(y) \, \Gamma \right] \otimes \Phi_{q \bar{q}}^{\Gamma} (y)
+
 {\rm tr} \left[ H^{\perp\mu}_{q \bar{q}}(y)  \Gamma \right] \otimes \Phi^{\perp\Gamma}_{{q \bar{q}}\,\mu} (y) + {\rm tr} \left[ H_{q \bar{q}g}^\mu(y_1,y_2) \, \Gamma \right] \otimes \Phi^{\Gamma}_{{q \bar{q}g}\,\mu} (y_1,y_2) \,,
\eea
% 
% 
% \bea
% \label{GenAmpFac23}
% \vspace{-.4cm}{\cal A}&=&
% \int\limits_{0}^{1} dy \ {\rm tr} \left[ H_{q \bar{q}}(y) \, \Gamma \right] \, \Phi_{q \bar{q}}^{\Gamma} (y)
% +
% \int\limits_{0}^{1} dy \ {\rm tr} \left[ H^{\perp\mu}_{q \bar{q}}(y) \, \Gamma \right] \, \Phi^{\perp\Gamma}_{{q \bar{q}}\,\mu} (y) \nonumber 
% \\
% &+&\int\limits_{0}^{1} dy_1\, dy_2 \ {\rm tr} \left[ H_{q \bar{q}g}^\mu(y_1,y_2) \, \Gamma \right] \, \Phi^{\Gamma}_{{q \bar{q}g}\,\mu} (y_1,y_2) \,,
% \eea
% in which
% the two first terms in the r.h.s corresponds to the 2-parton contribution and the last one to the 3-parton contribution (see Fig.\ref{Fig:Factorized2AND3body}).
where $\otimes$ is the $y$-integration.
Although the fields coordinates $z_i$ are on the light-cone in both LCCF and CCF parametrizations of the soft non-local correlators,
 $z_i$ is along $n$ in LCCF while arbitrary in CCF.
%(see, for example, (\ref{par1v})-(\ref{par1.1a}) and (\ref{cov}))
The transverse physical polarization of the $\rho-$meson is defined by the conditions
\beq
\label{pol_RhoTdef}
e_T \cdot n=e_T \cdot p=0\,.
\eq
Keeping all the terms up to the twist-$3$ order
with the axial (light-like) gauge, $n \cdot A=0$,
the matrix elements of quark-antiquark nonlocal operators
% and $\stackrel{{\cal F}_1}{=}$ is
% the Fourier transformation with measure
%$\int_{0}^{1}\, dy \,\text{exp}\left[iy\,p\cdot z\right]$
 for vector and axial-vector correlators without and with transverse derivatives,
with $\stackrel{\longleftrightarrow}
{\partial_{\rho}}=\frac{1}{2}(\stackrel{\longrightarrow}
{\partial_{\rho}}-\stackrel{\longleftarrow}{\partial_{\rho}})\,,$
can be written 
%respectively
%in terms of the light-cone basis vectors 
as (here, $z=\lambda n$)
\begin{eqnarray}
\label{par1v}
\langle \rho(p_\rho)|\bar\psi(z)\gamma_{\mu} \psi(0)|0\rangle
&=&
m_\rho\,f_\rho \int_{0}^{1}\, dy \, {\rm exp}\left[iy\,p\cdot z\right] \left[ \varphi_1(y)\, (e^*\cdot n)p_{\mu}+\varphi_3(y)\, e^*_{T\mu}\right],\,\,\,\,
\\
\label{par1.1v}
\langle \rho(p_\rho)|
\bar\psi(z)\gamma_{\mu}
i\stackrel{\longleftrightarrow}
{\partial^T_{\alpha}} \psi(0)|0 \rangle
&=& m_\rho\,f_\rho \,\int_{0}^{1}\, dy \, {\rm exp}\left[iy\,p\cdot z\right]
\varphi_1^T(y) \, p_{\mu} e^*_{T\alpha}\,, \\
%\end{eqnarray}
%and
%\begin{eqnarray}
% \label{par1v}
% &&\langle \rho(p_\rho)|\bar\psi(z)\gamma_{\mu} \psi(0)|0\rangle
% \stackrel{{\cal F}_1}{=}
% m_\rho\,f_\rho \left[ \varphi_1(y)\, (e^*\cdot n)p_{\mu}+\varphi_3(y)\, e^*_{T\mu}\right],
% \\
\label{par1a}
\langle \rho(p_\rho)|
\bar\psi(z)\gamma_5\gamma_{\mu} \psi(0) |0\rangle &=&
m_\rho\,f_\rho \, i\int_{0}^{1}\, dy \, {\rm exp}\left[iy\,p\cdot z\right]\varphi_A(y)\, \varepsilon_{\mu\alpha\beta\delta}\,
e^{*\alpha}_{T}p^{\beta}n^{\delta} \,,  \\
\label{par1.1a}
\langle \rho(p_\rho)| \bar\psi(z)\gamma_5\gamma_{\mu}
i\stackrel{\longleftrightarrow}
{\partial^T_{\alpha}} \psi(0) |0\rangle &=&
m_\rho\,f_\rho \,
i\int_{0}^{1}\, dy \, {\rm exp}\left[iy\,p\cdot z\right]\varphi_A^T (y) \, p_{\mu}\, \varepsilon_{\alpha\lambda\beta\delta}\,
e_T^{*\lambda} p^{\beta}\,n^{\delta}\,,
\end{eqnarray}
% where the corresponding flavour matrix has been omitted.
where 
$y$ ($\bar y$) is the quark (antiquark) momentum fraction.
Two analogous correlators are needed to describe gluonic degrees of freedom, introducing $B$ and $D$ DAs according to
\begin{eqnarray}
\label{Correlator3BodyV}
\hspace{-.8cm}\langle \rho(p_\rho)|
\bar\psi(z_1)\gamma_{\mu}g A_{\alpha}^T(z_2) \psi(0) |0\rangle \!
&=&\! m_\rho \,\fV \!
\int\limits_{0}^{1} \! dy_1 \! \int\limits_{0}^{1} \! dy_2 \,
%\text{exp}
e^{iy_1\,p\cdot z_1+i(y_2-y_1)\,p\cdot z_2 } \,
B(y_1,y_2) p_{\mu} e^*_{T\alpha}\,, \ \
\\
\label{Correlator3BodyA}
\hspace{-.6cm}\langle \rho(p_\rho)|
\bar\psi(z_1)\gamma_5\gamma_{\mu} g A_{\alpha}^T(z_2) \psi(0) |0\rangle
&=&m_\rho\,\fA \,
\int\limits_{0}^{1} dy_1 \,\int\limits_{0}^{1} dy_2 \,
e^{iy_1\,p\cdot z_1+i(y_2-y_1)\,p\cdot z_2 } \,
i D(y_1,y_2)\\
&&\hspace{1cm}\times p_{\mu} \, \varepsilon_{\alpha\lambda\beta\delta} \,
e^{* \, \lambda}_T \, p^{\beta}n^{\delta}\,.
\end{eqnarray}
 One thus needs 7 DAs:  $\varphi_1$ (twist-$2$), 
$B$ and $D$ (genuine (dynamical) twist-$3$) and
$\varphi_3$, $\varphi_A, \varphi_1^T$, $\varphi_A^T$
(kinematical (\`a la
Wandzura-Wilczek) twist-$3$ and genuine (dynamical) twist-$3$).

These DAs are not independent.
They are related by 2 Equations of Motions (EOMs) and 2 equations arising from the invariance of  ${\cal A}$ under the arbitrary vector
%rotation on the light-cone, i.e. based on the arbitrariness of 
$n,$
which  comes from 3 sources.
 First, it enters the definition
of the non-local correlators
 through
the light-like separation $z=\lambda \, n$.
%
% intering the non-local correlators
%introduced in Sec.\ref{SubSubSec_ParamVacuumRhoLCCF}
%involves an arbitrary light-like vector $n$ with $n \cdot p=1$
%which already appeared in the Sudakov decomposition (\ref{k}).
These correlators are defined in the axial light-like gauge $n \cdot A=0\,,$
which allows to get rid of  Wilson lines.
Second, it determines the notion of transverse polarization of the $\rho\,.$
Last, $n$ enters the Sudakov decomposition
(\ref{k}) which defines the transverse parton momentum involved in the collinear factorization. One can in fact show that the hard part does not depend on the gauge fixing vector $n.$ Therefore, only the second and third source of $n-$dependence should be investigated.  Based on Ward identities, this $n-$dependence of 
${\cal A}$ can be recast 
in a system of constraints which only
involve  
%Indeed, this invariance  with respect to $n$ does not involve the hard part of ${\cal A}$,
%in  
%and therefore implies constraints on 
the soft part.
%, i.e. on the DAs. 
We thus have only 3 independent DAs $\varphi_1$ , 
$B$ and $D$, which fully encode  \pagebreak

\noindent 
the non-perturbative content of the $\rho$ at twist 3.

 The original CCF parametrizations of the $\rho$ DAs~\cite{BB} also  involve 3 independent DAs, defined through 4  correlators related by EOMs. The 2-parton axial-vector correlator reads, 
\beq
\label{BBA}
\langle \rho(p_\rho)|\bar \psi(z) \, [z,\, 0] \, \gamma_\mu \gamma_5 \psi(0)|0\rangle =
\frac{1}{4}f_\rho\,m_\rho\, \varepsilon_\mu^{\,\,\,\alpha \beta \gamma} e^*_{T \alpha} \,p_\beta \, z_\gamma\,    \int\limits_0^1\,dy\,e^{iy(p \cdot z)}\,g_\perp^{(a)}(y)\;,
\eeq
%
%\beq
%\label{defWilson}
$[z_1, \, z_2] = P \exp \left[ i g \int\limits^1_0 dt \, (z_1-z_2)_\mu A^\mu(t \,z_1 +(1-t)\,z_2    \right]$
%\eq
being the Wilson line. Denoting the meson polarization vector by $e,$
 $e_T$ is here defined to be orthogonal to the light-cone vectors $p$ and $z$:
\beq
\label{pol_Rho}
e_{T \mu}=e_\mu -p_\mu \frac{e \cdot z}{p \cdot z}-z_\mu \frac{e \cdot p}{p \cdot z} \, .
\eeq
Thus  $e_T$  (\ref{pol_Rho}) in CCF and $e_T$ (\ref{pol_RhoTdef}) in LCCF differ since  $z$ does not generally point in the $n$ direction.
The  2-parton vector correlator reads (up to twist 3)
\beq
\label{BBV1}
\langle \rho(p_\rho)|\bar \psi(z) \, [z,\, 0] \, \gamma_\mu  \psi(0)|0\rangle = f_\rho\,m_\rho\int\limits_0^1\,dy\,e^{iy(p\cdot z)}\left[
p_\mu\,\frac{e^*\cdot z}{p\cdot z}\phi_{\parallel}(y) +
e^*_{T\mu}\,g_\perp^{(v)}(y) 
%-z_\mu \frac{m^2}{2} \frac{e^* \cdot z}{(p \cdot z)^2}  g_3(y) 
\right]
\,.
\eq
The 3-parton correlators are parametrized 
 (up to twist 3 level) according to
\bea
\hspace{-1cm}\langle \rho(p_\rho)|\bar \psi(z)[z,t\, z]\gamma_\alpha g \, G_{\mu\nu}(t\, z)[t\,z,0] \psi(0)|0 \rangle &=&
-i p_\alpha [p_\mu e^*_{\perp \nu}-p_\nu e^*_{\perp \mu} ] \, m_\rho \, \fV \nonumber \\
&\times& \int D \alpha \, V(\alpha_1,\alpha_2) \,
e^{\,i p \cdot z \,(\alpha_1+\,t\,\alpha_g)} \, , \label{GV}\\
\hspace{-1cm}\langle \rho(p_\rho)|\bar \psi(z)[z,t\, z]\gamma_\alpha\gamma_5 g \, \tilde G_{\mu\nu}(t\, z)[t\,z,0] \psi(0)|0 \rangle &=&
- p_\alpha [p_\mu e^*_{\perp \nu}-p_\nu e^*_{\perp \mu} ] \, m_\rho \,\fA \nonumber \\
&\times & \int D \alpha \, A(\alpha_1,\alpha_2) \,
e^{\,i \, p \cdot z \,(\alpha_1+\,t\,\alpha_g)} \,,\label{GA}
\eea
where $\alpha_1$, $\alpha_2$, $\alpha_g$ are momentum fractions of quark, antiquark and gluon respectively inside the $\rho-$meson, 
%\beq
%\label{defDalpha}
$\int D \alpha =\int\limits^1_0 d\alpha_1\int\limits^1_0 d\alpha_2 \int\limits^1_0 d\alpha_g\,
\delta(1-\alpha_1-\alpha_2-\alpha_g)$
%\eq
and $\tilde G_{\mu\nu}=-{1\over 2}\epsilon_{\mu\nu\alpha\beta}G^{\alpha\beta}.$ 
A comparison of the  correlators (\ref{par1v}, \ref{par1.1v}, \ref{par1a}, \ref{par1.1a}, \ref{Correlator3BodyV}, \ref{Correlator3BodyA}) and (\ref{BBA}, \ref{BBV1}, \ref{GV}, \ref{GA})  in the axial gauge $n \cdot A=0$ gives the following
identification of the 2- and 3-parton DAs in LCCF and CCF  approaches:
\begin{eqnarray}
\label{relBBvector-axial}
&&\varphi_1(y)=
\phi_{\parallel}(y) ,
\quad
\varphi_3(y)=
 g_\perp^{(v)}(y) \,, \quad
%\label{relBBaxial}
\varphi_A(y) =
-\frac{1}{4} \, \frac{\partial g_\perp^{(a)}(y)}{\partial y}\,,
\\
\label{DictB-D}
 &&B(y_1,\,y_2)=-\frac{V(y_1, \, 1-y_2)}{y_2-y_1}\, \quad
%\\
%\label{DictD}
D(y_1,\,y_2)=-\frac{A(y_1, \, 1-y_2)}{y_2-y_1}\,.
\end{eqnarray}

% We thus use Eq.(\ref{pol_Rho}) and rewrite the original CCF 
% parametrization in terms of the full meson polarization vector $e$. This is already done for the axial-vector correlator (\ref{BBA})
%  since due to the properties of fully antisymmetric tensor 
% $\epsilon_{\mu\nu\alpha\beta}$ one can use in the r.h.s. of (\ref{BBA}) the full meson polarization vector
% $e$ instead of $e_T$. The same treatment can be done for the 2-parton vector correlator.

%\section{A minimal set of chiral-event Distribution Amplitudes for $\rho$ at twist three}

\section{$\gamma^* \to \rho_T$ Impact factor up to  twist three accuracy in LCCF and CCF}

% The forward impact factor %$\Phi^{\gamma^*\to\rho}(\kb,\,-\kb)$ 
% of the subprocess
%  $g(k_1,\varepsilon_{1})+\gamma^*(q)\to g(k_2, \varepsilon_{2})+\rho_T(p_1)$
% is
%  defined as
% the integral of the  discontinuity in the $g_1+\gamma^*$ channel of the  off-shell S-matrix element 
%  ${\cal S}^{\gamma^*_T\, g\to\rho_T\, g}_\mu$.
%  We calculated $\Phi^{\gamma^*\to\rho}$ using both LCCF and CCF approaches.
%%%%%
We have calculated, in both LCCF and CCF, the forward impact factor $\Phi^{\gamma^*\to\rho}$ %$\Phi^{\gamma^*\to\rho}(\kb,\,-\kb)$ 
of the subprocess
 $g+\gamma^*\to g+\rho_T\,,$
 defined as
the integral of the  discontinuity in the $s$
%$g_1+\gamma^*$ 
channel of the  off-shell S-matrix element 
 ${\cal S}^{\gamma^*_T\, g\to\rho_T\, g}_\mu$.
  In LCCF, one computes the diagrams perturbatively in a fairly direct way,
which makes the use of the  CCF
 parametrization \cite{BB} less practical.
 We need to express the impact factor in terms of 
hard coefficient functions and soft parts parametrized by the light-cone  matrix
 elements. The standard technique 
here is an operator product expansion on the light cone, which
gives the leading term in the power counting.
 Since there is no operator definition for \linebreak an  impact
factor, we have to rely on perturbation theory. The primary complication encountered is
% and leads to the described above factorized structure. 
\pagebreak
 
\noindent
 %in our actual calculation 
  that the $z^2\to 0$ limit of any single diagram is given in terms 
of  
light-cone 
matrix elements without any %\linebreak    
Wilson
line insertion between the quark and gluon operators (''perturbative correlators``), like
$
\langle \rho(p_\rho)|\bar \psi(z)\gamma_\mu \psi(0)|0 \rangle\,.$
% \quad {\rm and}\quad  \langle V(p_V)|\bar \psi(z)\gamma_\mu A_\alpha (t\, z) \psi(0)|0 \rangle \, ,
%$
%which we call conventionally  ''perturbative correlators``. 
%Actually we need to combine together contributions of quark-antiquark and quark-antiquark gluon diagrams in order to obtain final gauge invariant result.
Despite working in the axial gauge one cannot neglect  effects coming from the Wilson
lines since the  two light cone vectors $z$ and $n$ are not identical and thus, generically, Wilson lines are not equal to unity. Nevertheless in the axial gauge the contribution of each additional parton costs one extra power 
of $1/Q$, allowing the calculation to be
organized in a simple iterative manner expanding the Wilson line. 
At twist 3, we need to keep the contribution
%three level it is enough to consider the first two terms of such expansion 
%\beq
%\label{wl}
$[z,0]=1+i \,g \int\limits^1_0 dt \, z^\alpha A_\alpha (z t)$ and to care about the difference between  the physical  $\rho_T$-polarization (\ref{pol_RhoTdef}) from the formal one  (\ref{pol_Rho}).
At twist 3-level the net effect of the Wilson line when computing our impact factor is
just a renormalization of the DA  $g^a_\perp$  of  (\ref{BBA}), and similarly for the vector case.

%. A similar result is obtained for the vector case. 

Based %on the dictionary which we established between LCCF and CCF, and
 on the solution of the EOMs and $n$-independence set of equations,  our two LCCF and CCF results are identical; they are gauge invariant due to a consistent inclusion of fermionic and gluonic degrees of freedom and  are free of end-point singularities, due to the $k_T$ regulator. 
\vspace{.3cm}

%  In conclusion, we have proven the equivalence of LCCF and CCF approaches at twist 3, given the dictionary, and checked this correspondence in the case of the  $\gamma^* \to \rho_T$ impact factor.

 This work is partly supported by  the grant
ANR-06-JCJC-0084, the RFBR (grants 09-02-01149,
 08-02-00334, 08-02-00896), the grant 
NSh-3810.2010.2 
and
the Polish Grant N202 249235.

\end{document}